\documentclass[utf8]{FrontiersinHarvard} 

\usepackage{url,hyperref,microtype,subcaption}
\usepackage[onehalfspacing]{setspace}

\def\keyFont{\fontsize{8}{11}\helveticabold }
\def\firstAuthorLast{Brooks {et~al.}} 
\def\Authors{David H. Brooks\,$^{1,*}$, Jeffrey W. Reep\,$^{2}$, Ignacio Ugarte-Urra\,$^{2}$ and Harry P. Warren\,$^{2}$}
\def\Address{$^{1}$College of Science, George Mason University, 4400 University Drive, Fairfax, VA 22030, USA\\
$^{2}$Space Science Division, Naval Research Laboratory, Washington, DC 20375, USA  }
\def\corrAuthor{David H. Brooks}
\def\corrEmail{dhbrooks.work@gmail.com}

\begin{document}
\onecolumn
\firstpage{1}

\title[Hinode/EIS flare trigger]{On orbit performance of the solar flare trigger for the Hinode EUV Imaging Spectrometer} 

\author[\firstAuthorLast ]{\Authors} 
\address{} 
\correspondance{} 

\extraAuth{}

\maketitle

\begin{abstract}
We assess the on-orbit performance of the flare event trigger for the Hinode EUV Imaging Spectrometer. Our goal is to understand the time-delay between
the occurrence of a flare, as defined by a prompt rise in soft X-ray emission, and the initiation of the response observing study. Wide (266$''$) slit
patrol images in the He II 256.32\,\AA\, spectral line are used for flare hunting, and a reponse is triggered when a pre-defined intensity threshold
is reached. We use a sample of 13 $>$ M-class flares that succesfully triggered a response, and compare the timings with soft X-ray data from GOES, 
and hard X-ray data from RHESSI and Fermi. Excluding complex events that are difficult to interpret, the mean on orbit response time for our sample is 2 min
10 s, with an uncertainty of 84 s. These results may be useful for planning autonomous operations for future missions, and give some guidance as to how improvements could be made
to capture the important impulsive phase of flares.

\section{}

\tiny
 \keyFont{ \section{Keywords:} Solar Flares, Solar Instrumentation} 
\end{abstract}

\section{Introduction}
A central goal of heliophysics research is to understand the build-up, storage, and release of magnetic energy during solar flares and coronal
mass ejections (CMEs). As well as scientific understanding of the phenomena themselves, another eventual goal is to develop the ability to 
predict their occurrence, and mitigate the effects of CME and energetic particle impact on the terrestrial space environment. 

High spatial and temporal resolution extreme ultraviolet (EUV) imaging of almost all flares on the Earth-facing side of the Sun is possible
with full-disk EUV instruments such as the Atmospheric Imaging Assembly \citep[AIA,][]{Lemen2012} on the Solar Dynamics Observatory \citep[SDO,][]{Pesnell2012}.
For deeper scientific insights, however, spectroscopic diagnostic measurements are highly desirable. The drawback is that current EUV spectrometer
fields-of-view (FOV) are limited (typically covering something like $\sim$250$''$x500$''$) and the slit rastering times are 
slow (typical timescales of hours), so such observations are not ideal for capturing dynamic events. Full-disk imaging with slit spectrometers 
is possible and has occasionally been carried out \citep{Thompson2000,Brooks2015,Bryans2020}, but the scanning times are even slower.
Routine full-disk scans by the Hinode EUV Imaging Spectrometer \citep[EIS,][]{Culhane2007}, for example,
take about 40 hours to complete. Proposed future missions for full-disk spectroscopy could improve this situation \citep{Ugarte2023}, but for
the near term, spectroscopic instruments in development, such as Solar-C EUVST and MUSE \citep[MUlti-slit Solar Explorer,][]{DePontieu2020}, will be limited to relatively small fields-of-view.

The scientific value of these observations is high. EIS, for example, has a wide range of spectral diagnostics that allow measurements of temperatures, densities, 
elemental abundances, Doppler and non-thermal velocities, and coronal magnetic field strengths in flares \citep{Doschek2018,Warren2018,To2021,Landi2021}. For reviews
of flare observations by EIS see, e.g., \cite{Milligan2015} and \cite{Hinode2019}. The downside is
that the small FOV and slow scanning times make it difficult to capture flares. Together with the difficulty in predicting which active regions are likely to
flare, especially when there are multiple possible targets on disk, and restrictions on the quantities of data that can be telemetered to ground, it becomes
challenging to maximise the scientific output.

There have been some studies of the success rate of flare observations for instruments such as EIS. \cite{Inglis2021} simulated the peformance of 
a small FOV mission depending on different operational and pointing selection strategies. They found that the most successful strategy for capturing
flares was based on response to actual flaring activity in the previous 24 hours. Depending on whether the satellite was placed in low Earth orbit, or 
had continuous solar observing coverage, 35--62\% of M- and X-class flares could be captured with our current forecasting abilities and
observing strategies. An important result is that they also found that the success rate was highly dependent on the
delay time between acquiring target information and re-pointing the instrument/spacecraft. A quick response is vital.

\cite{Watanabe2012} found that for the first 5 years
of the Hinode mission, EIS was able to observe on the order of 15\% of flares that reached C-class, 
as defined by the Geostationary Operational Environmental Satellites (GOES). The percentage is higher (20\%) for M- and X-class flares. These figures, however,
do not take account of the fact that in periods of high solar activity there can be multiple flaring regions on disk at the same time and, as discussed, EIS cannot
observe all of them because of its small FOV. In that sense, the numbers do not necessarily reflect how good the target selection strategy was, since the presence of 
several flaring ARs can dilute the numbers. The rate has 
also evolved as a result of improved strategies for flare 
observing during the mission, but it should be noted that lower telemetry observing studies that can run for a long duration are likely more successful
at capturing flares than higher cadence studies with more diagnostics that can only be used for shorter periods. 

An alternative to routine observing modes that monitor for flares is to implement an on board autonomous event trigger that responds to flare occurrence. 
In this way a low cadence flare monitoring program can switch to a higher cadence study when an event is detected. For EIS the data volume for download each
day is on the order of $\sim$ 750\,Mb. A typical flare response study that is currently used consumes 270\,Mb per hour, so it is clear that the trigger can make better use of the
available telemetry. Furthermore, a quick response to a flare potentially allows observations of the important impulsive phase \citep{Harra2009,Jeffrey2018}. Note that the currently available data volume for EIS is about 15--20\% of the original baseline as a result of an issue with the X-band
antenna in 2007 \citep{Hinode2019}. This limitation was somewhat mitigated by an increase in the number of downlink stations worldwide, and 
the introduction of more flexible control of data acquisition.

For EIS the pre-flight scientific requirement was to react and initiate a response program within 30\,s of flare detection. This requirement is met by
the on-board software: the mean time to start the response study from the trigger time in our sample of observations (see below) is 17.2\,s. This does not, however,
tell us how quickly the instrument responds in practice to the start of the actual flare, 
information that is important for the development of future missions. Here we
investigate the response times for a sample of large flares captured by the EIS flare trigger on orbit and report the results.

\section{Method}
Hinode/EIS is described in detail by \cite{Culhane2007}. The instrument is a normal incidence spectrograph that observes in two 
wavelength ranges from 171--211\,\AA\, and 245--291\,\AA\, with a spectral resolution of 23\,m\AA. A rotating slit assembly allows
the use of four different apertures of 1$''$, 2$''$, 40$''$, and 266$''$ widths. EIS has two internal and one external solar event triggers. 
Internally there is a bright point trigger. This locates an area of 
maximum (pre-defined) intensity within a narrow slit (1$''$ or 2$''$) raster scan and its functionality is discussed in EIS software note No. 14 by 
Young (2011)\footnote{\url{http://solarb.mssl.ucl.ac.uk/SolarB/eis_docs/eis_notes/14_BP_TRIGGER/eis_swnote_14.pdf}}.
A key point is that the raster scan is completed before the response study begins, so one drawback is that the detection is slow. 
Hence this trigger is not suitable for determining how quickly EIS responds to
the appearance of a bright point. It is also not suitable for observing flares.    
Externally, EIS can respond to a flare flag raised by Hinode/XRT \citep[X-ray Telescope,][]{golub2007}. In this case, EIS takes patrol images
purely to switch into flare response mode (one image is enough). There is no relationship between the patrol images and the trigger for the
flare. These two event triggers are not discussed further here.
EIS also has its own internal flare trigger (EFT). In this brief report we focus on the performance of this event trigger.

For flare hunting with the EFT, the 266$''$ wide-slit is used to monitor a typical FOV of 266$''$x512$''$.
Typically the He II 256.32\,\AA\, spectral line is selected for active region monitoring, though 
experiments with Fe XXIV 192.04\,\AA\, have also been undertaken. The advantage of He II is that the upper chromosphere often brightens first
(in response to the impact of non-thermal electrons) before post-flare loops begin to be filled with higher temperature emitting plasma 
(that would trigger in Fe XXIV or soft X-rays). We note that opacity effects that radiatively couple different points within the 
emitting plasma can affect the emergent intensities of He II 256.32\,\AA. Whether this has any effect on the speed of production of an
observable flare signature is an open question. It could be that another optically thin, or weakly modified optically thick, chromospheric
or transition region line would lead to an improved trigger performance.

A detection algorithm is run on each exposure. This compares the He II 256.32\,\AA\, intensity distributions in the X- and Y- directions on the detector
with several pre-defined criteria. First, the intensities are summed along each row and column in the image. Note that the column intensities correspond 
to the solar-Y direction. Because the wide-slit is used,
however, the spatial and spectral information is convolved in the dispersion direction and the row intensities correspond to positions in solar-X and
wavelength. Second, the summed row and column intensity distributions are compared to pre-defined thresholds, and the number of pixels that exceed
the thresholds in both directions are recorded. Third, a check is made to determine how many of the excess pixels occur consecutively
in each direction. If these numbers exceed another pre-defined threshold then the flare response is triggered. We show an illustration of the
functionality of the algorithm in the next section.

All of the thresholds can be
adjusted by the EIS Chief Observer, but in practice care has to be taken so experiments
were performed to find the best values and these are not routinely changed.
For example, in the Y-direction, the
EIS slit is long, so the active region contribution to the summed intensity can be relatively small. The threshold in this direction is therefore set
relatively low so that the detection criteria are almost always satisfied by a flare. On-orbit testing determined that a value of 350,000\,DN\, successfully
triggers. In the X-direction, the threshold is set higher so that only large
flares are detected. Again, on-orbit testing determined that a value of 400,000\,DN\, is high enough to exclude events weaker than around M-class. Currently,
the same thresholds are used for all EFT runs. The consecutive-pixels lower limit is implemented to avoid triggering on energetic particle hits. Hinode passes through the
South Atlantic Anomaly (SAA) and High-Latitude Zone (HLZ) several times per day, but energetic particles tend to light up only a few pixels, or streak
across the detector so that they are not clustered consecutively. Although it is difficult to assess the history of the whole mission, we are not aware
of any occasions when the flare trigger was activated by SAA passage.

Once the flare occurrence criteria are satisfied,
the flare coordinates are determined. This can be done in two ways. The row and column summed intensities are examined to find either the location of the peak intensity in the X- and Y-directions, or the center pixel between the first and last crossing of the threshold (in both directions). The EIS Chief Observer can choose which method is used, but in practice the peak is normally used. EIS then re-points to the flare location using its fine mirror, and runs a pre-defined response study, typically
with a greater selection of diagnostic lines, higher cadence, shorter exposures (to avoid saturation), and higher telemetry output.

To compare the EFT detection times with the start of a flare we analyzed soft X-ray (SXR) and hard X-ray (HXR) observations from 
GOES, the Reuven Ramaty High Energy Solar Spectroscopic Imager \citep[RHESSI,][]{Lin2002} and the Fermi Gamma-ray Burst Monitor \citep[GBM,][]{Meegan2009}.
The GOES data were downloaded using the GOES workbench available in SolarSoftware \citep[SSW,][]{Freeland1998}. This software returns high cadence (2\,s) data 
in the 1.0--8.0\,\AA\, channel from the GOES satellite operating during the selected time-period. The data are archived at the
NOAA (National Oceanic and Atmospheric Administration) Space Weather Prediction Center (SWPC) in Boulder, CO. For RHESSI and Fermi, we used data from the 20--50\,keV\,
energy channel.
These data were retrieved from using the OSPEX (Objective Spectral Executive) software in SSW \citep{Tolbert2020}. The Fermi/GBM data were obtained from the NASA host website and the RHESSI data were transferred from the mirror site at the University of California, Berkeley. All the EIS data were processed using the standard (eis\_prep) calibration routines available in SSW.

We analyzed a sample of 13 $>$ M-class flares that were successfully detected by the EFT. 
We used the Hinode Flare Catalog \citep{Watanabe2012} to find M- and X-class
flares that were observed by EIS, and cross-checked which ones were captured by the EIS flare response study. For
all these observations the He II 256.32\,\AA\, spectral line was used for flare hunting. Examining the response time requires a definition of the flare start time.
The cataloged NOAA GOES start times are defined as a steep monotonic increase in the 1.0--8.0\,\AA\, flux in a sequence of 4\,min, but they use 1--5\,min average data
and take the first minute as the onset time. 
For an EFT reaction time measured in seconds, we need a start time defined to a similar time-resolution. We therefore follow their approach and define the
start of the flare as the time when the GOES SXR flux
increases promptly above the background, but using the high cadence (2\,s) data. Furthermore, we adopt a flexible definition of the prompt increase, since our sample
of flares is small enough for us to examine their characteristics individually.
Specifically, we take a 4\,min running mean of the background, and record when the SXR flux increases by 30\%. In a few
cases we lowered the threshold to 20\% or increased it to 50\%. This variable threshold is necessary because although most of the flares are large, impulsive
events, a few of them occur in the decaying tail of a previous flare so that the background is already high. The EIS trigger times were taken from the time
of the exposure that met the trigger criteria. 
\begin{figure}[h!]
\begin{center}
\includegraphics[width=18cm]{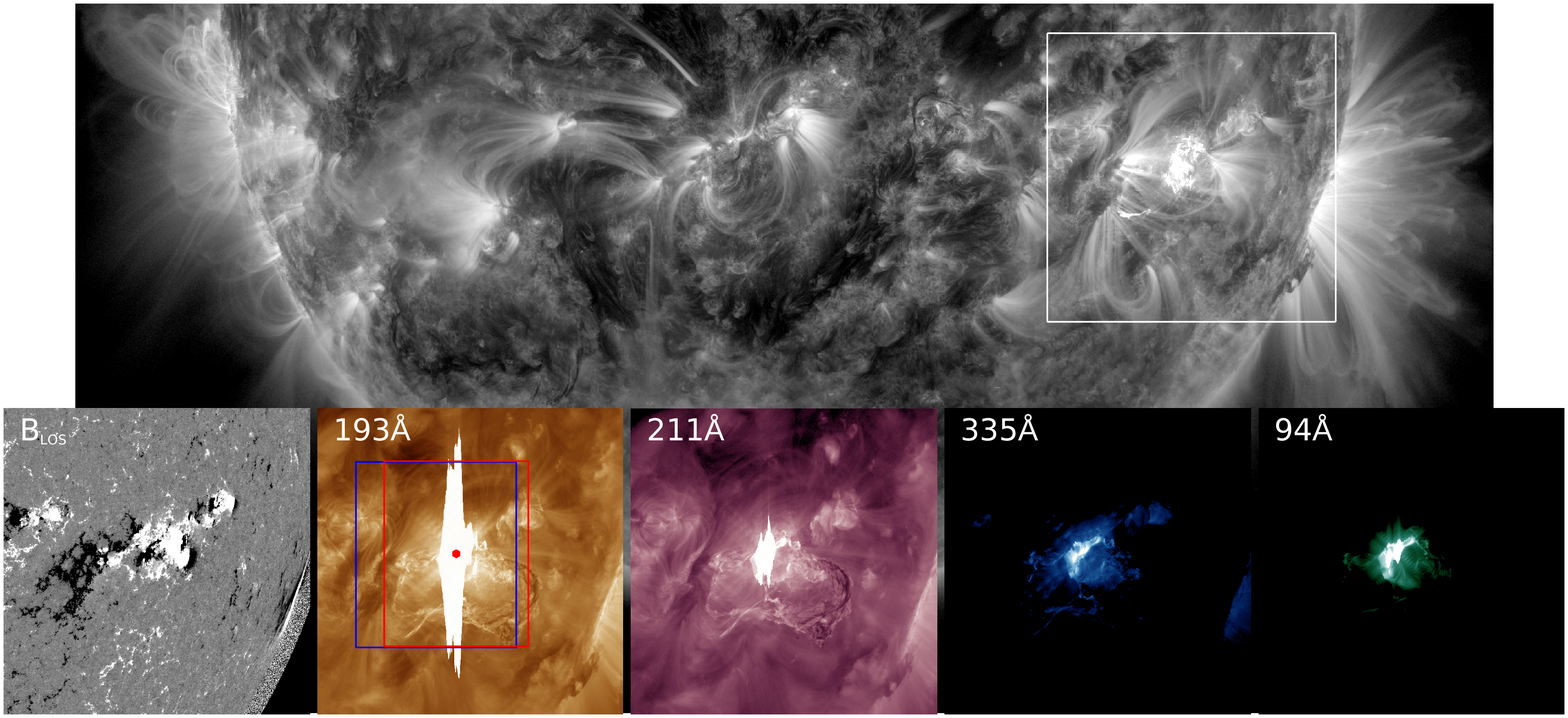}
\includegraphics[viewport = 0 0 1024 206,scale=0.50]{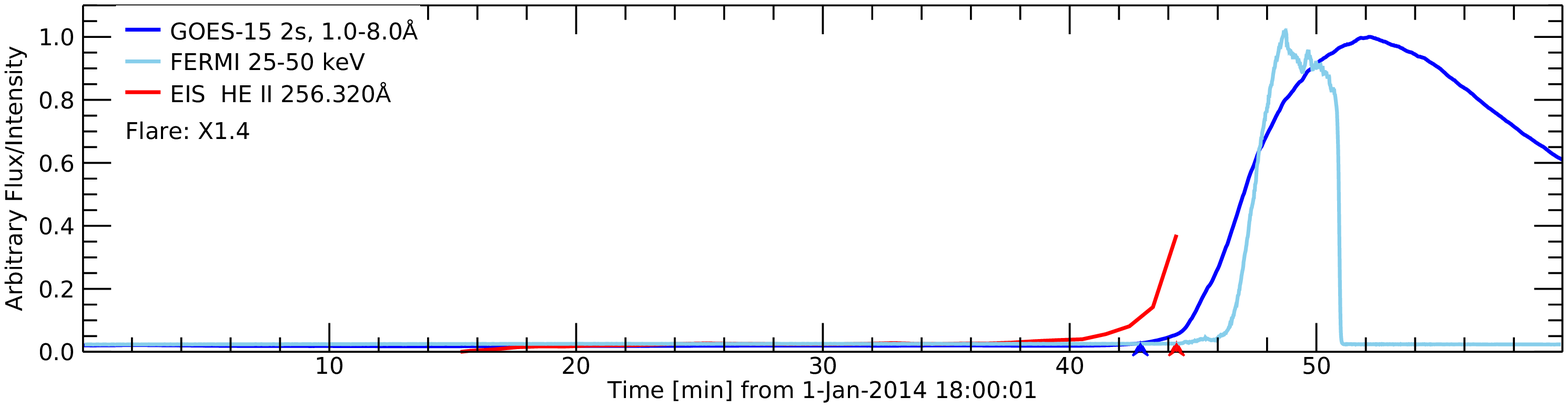}
\end{center}
\caption{ An example flare from our sample. The grey scale background is an AIA 171\,\AA\, image taken
at 18:44:32\,UT on 2014, January 1. The flare host AR 11936 is highlighted by the white box and the FOV
corresponds to the cutouts in the middle row showing (left to right) an HMI line-of-sight magnetogram,
AIA 193\,\AA, 211\,\AA, 335\,\AA, and 94\,\AA\, images. The AIA images are dominated by Fe lines forming
at temperatures of 1.4\,MK, 1.8\,MK, 2.5\,MK and 7.1\,MK\, in flaring conditions, respectively. The blue box shows the FOV of
the EIS wide-slit flare hunter study. Once triggered, EIS switches to the designated response study
centered on the detected flare site (red dot) with a different FOV (red box). The bottom row
shows the GOES soft X-ray (SXR) light curve (blue), Fermi hard X-ray (HXR) light curve (sky blue), and
EIS He II 256.32\,\AA\, mean intensity of the FOV. The abrupt end of the red line indicates when EIS was triggered. This
is also indicated by the red arrow. The blue arrow shows the start time of the flare derived from the SXR data (see text).
The X-axis labels give the time in minutes from the start time shown in the axis title. 
\label{fig:1}}
\end{figure}
The algorithm, of course, checks after the exposure is made, so there can be some latency that depends on the exposure
time. For our sample, the exposure time is 5\,s.

\section{Results}

We show an example of one of the flares in our sample in Figure \ref{fig:1}. The X1.4 flare occurred in AR 11936 on 2014, January 1.
The HMI magnetogram shows that AR 11936 had a magnetically complex $\beta\gamma\delta$ configuration. The GOES SXR flux shows that the
flare began at 18:42:52\,UT, with a steep rise in the Fermi HXR emission a couple of minutes later. EIS was running its flare hunter study 
from around 18:10\,UT\, and the He II 256.32\,\AA\, intensity began to increase from around 18:42\,UT. The lower panel plot gives 
\begin{figure}[h!]
\begin{center}
\includegraphics[viewport = 100 100 1848 598,width=18cm]{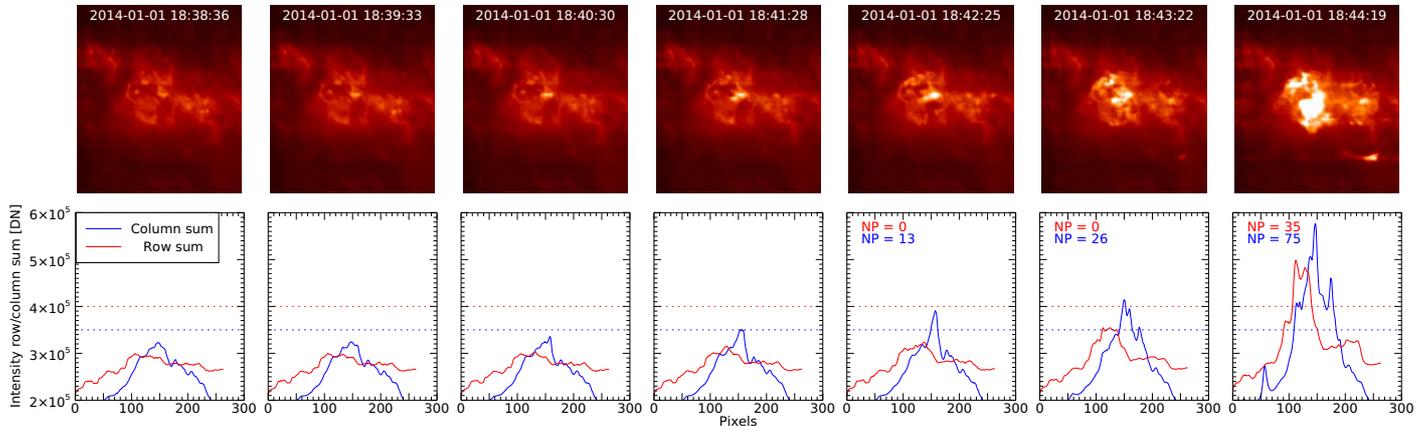}
\end{center}
\caption{ Illustration of the functionality of the EFT detection algorithm for the flare shown in Figure \ref{fig:1}. The
top row shows EIS He II 256.32\,\AA\, wide-slit hunter images in the 6\,min leading up to the flare detection. The bottom
row shows how the algorithm is evaluating the intensity distributions in the images. The solid red/blue lines show the 
intensities summed along the rows/columns. The dashed blue/red line show the detection thresholds. 
The number of points above the detection threshold are given in the legend. The X-axis labels show spatial pixels in 
the solar-X direction. 
\label{fig:2}}
\end{figure}
the impression
that EIS triggered earlier than the SXR rise, but of course the intensity has to reach the 
detection threshold before the instrument responds.
The detection is actually at 18:44:19\,UT, 87\,s\, after the SXR flare start time. 

Figure \ref{fig:2} shows the He II 256.32\,\AA\, intensity images in the 6\,min leading up to the flare. 
Note that the first 87\,s\, of the flare are not completely missed. Two patrol
images are taken in He II 256.32\,\AA. Even these images can be used to potentially deconvolve velocity information on the flare onset \citep{Harra2017}. 
There is a conspicuous pre-flare brightening, however, that is seen even earlier than the GOES flare start time, and therefore might be the ideal
trigger. The lower row in the figure illustrates
what is happening with the flare detection algorithm during this sequence of images. As discussed in the previous section,
the row sum threshold is set to 400,000\,DN, and the column sum threshold is set to 350,000\,DN, as is the case for most current runs. 
Neither intensity distribution exceeds the threshold until just before the GOES start time (5th frame at 18:42:25\,UT). At this time more than 13 consecutive
pixels exceed the threshold in the Y-direction. That is, the lower colum threshold has triggered. The row sum threshold, however, is not exceeded until 18:44:19\,UT, although it is close in the previous frame. By this time, 
all the threshold criteria are met, including the 7 consecutive-pixel limit, and the flare triggers. We can conclude that the trigger could have detected the flare one frame earlier if a
lower row sum threshold were used. The pre-flare brightening looks detectable even earlier in the column sum intensity distribution if a lower intensity
threshold were used, but it perhaps looks difficult to detect at all in the row summed intensities. The success of the detection criteria are of course
flare dependent. Note that using the peak intensities locates the flare at a pixel coordinate of [112,147].

\begin{figure*}[t!]
\centerline{%
\includegraphics[viewport = 20 10 610 490,scale=0.38]{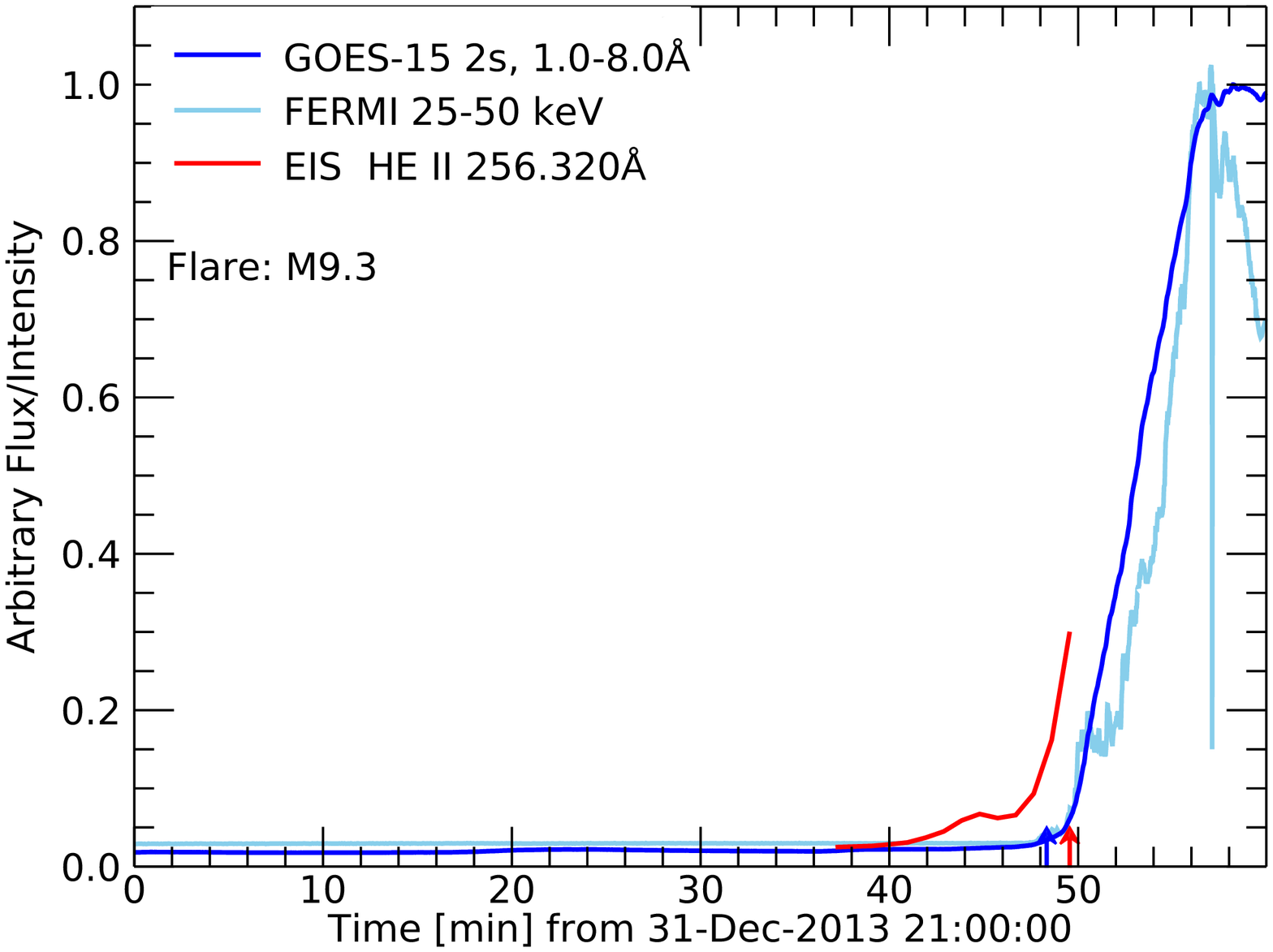}
\includegraphics[viewport = 20 10 610 490,scale=0.38]{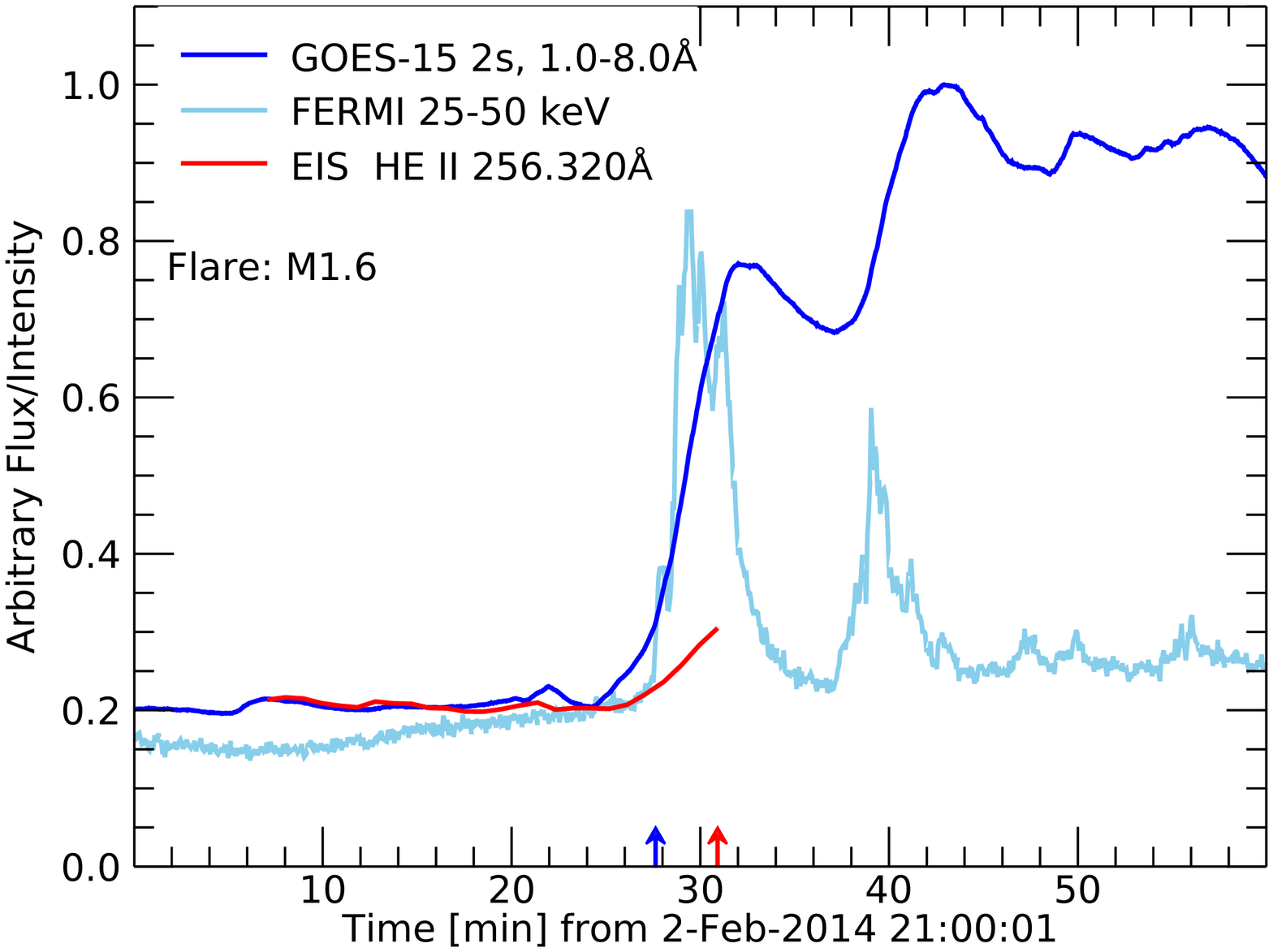}}
\centerline{%
\includegraphics[viewport = 20 10 610 490,scale=0.38]{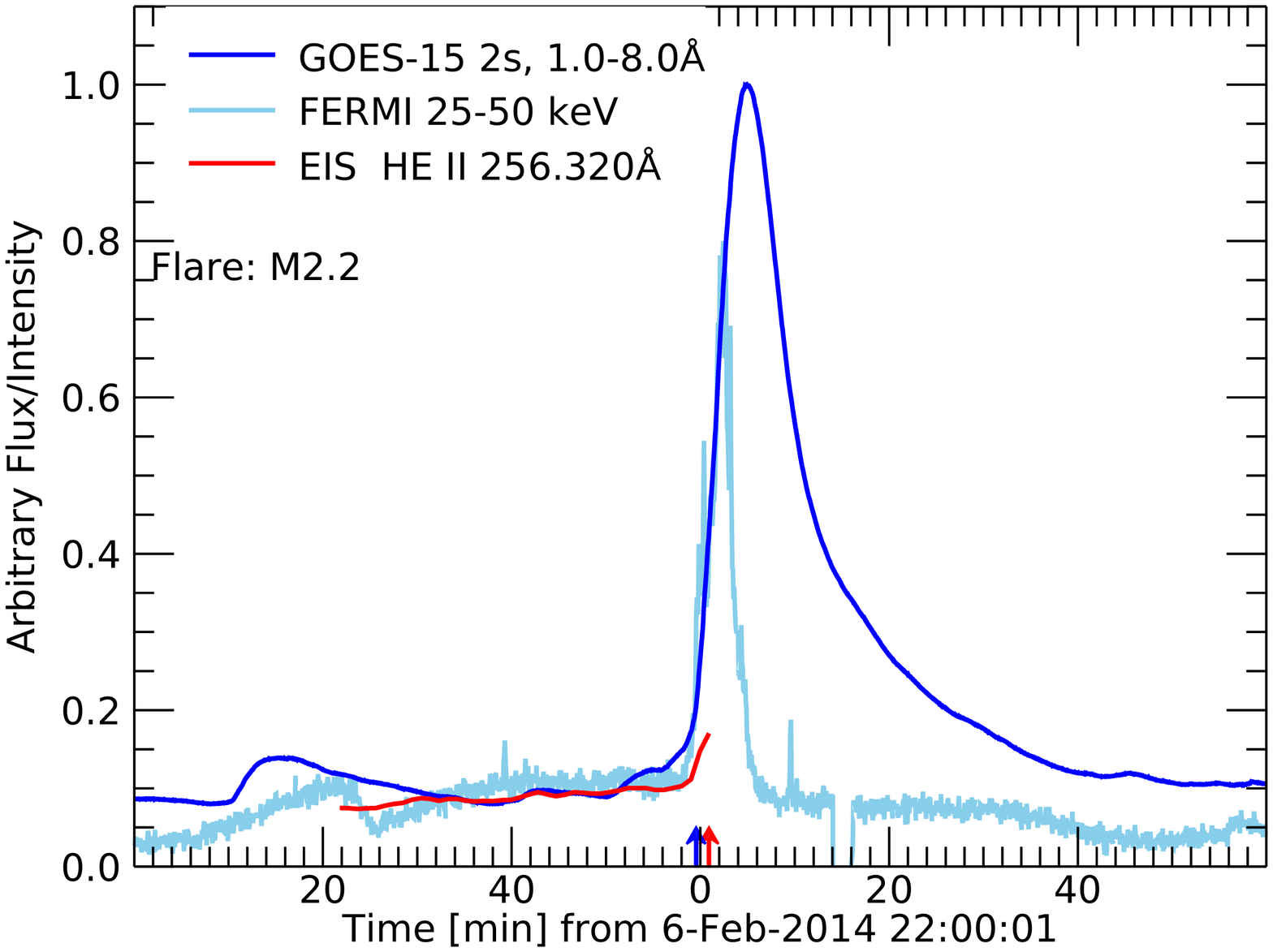}
\includegraphics[viewport = 20 10 610 490,scale=0.38]{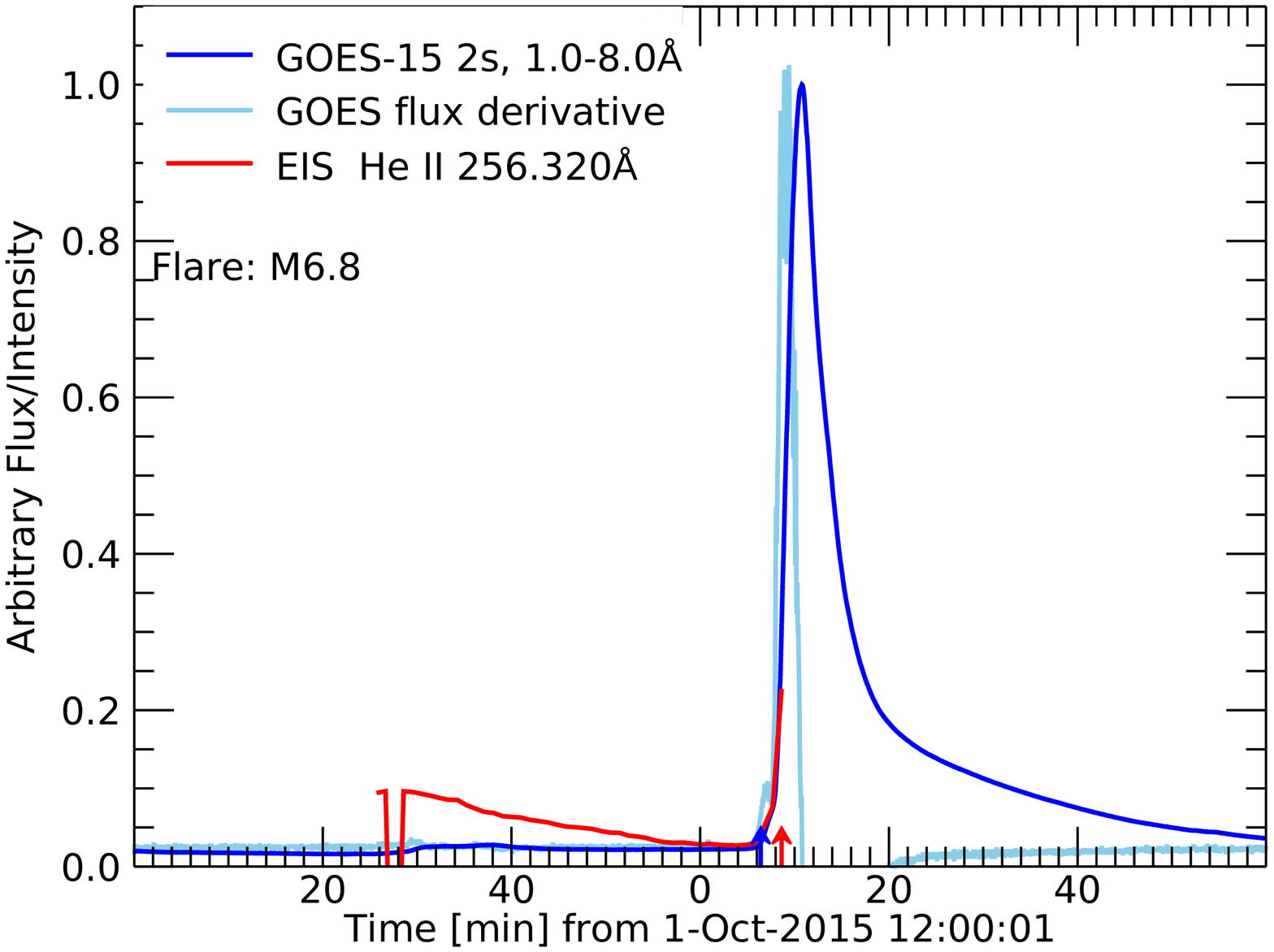}}
\centerline{%
\includegraphics[viewport = 20 10 610 490,scale=0.38]{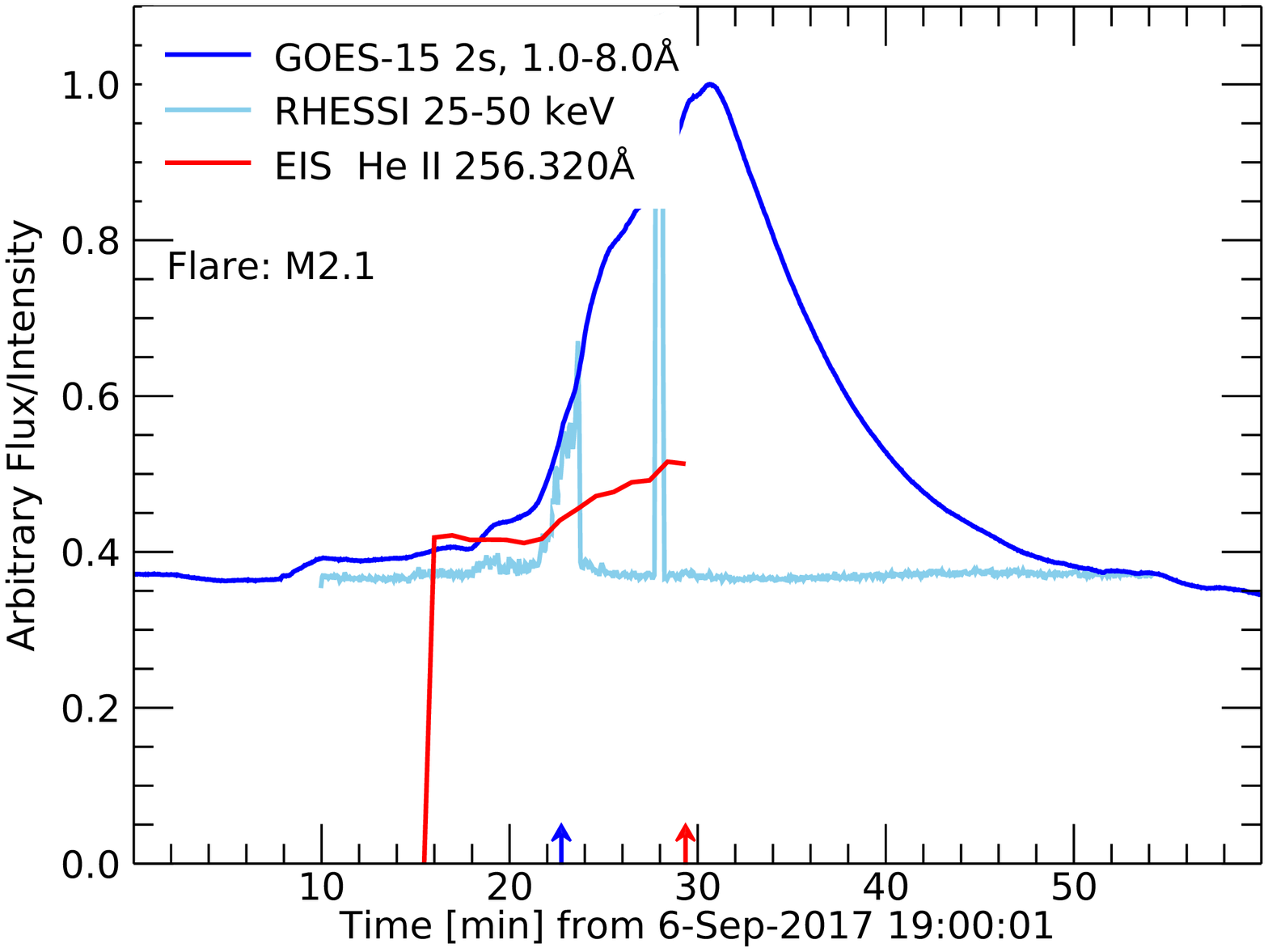}
\includegraphics[viewport = 20 10 610 490,scale=0.38]{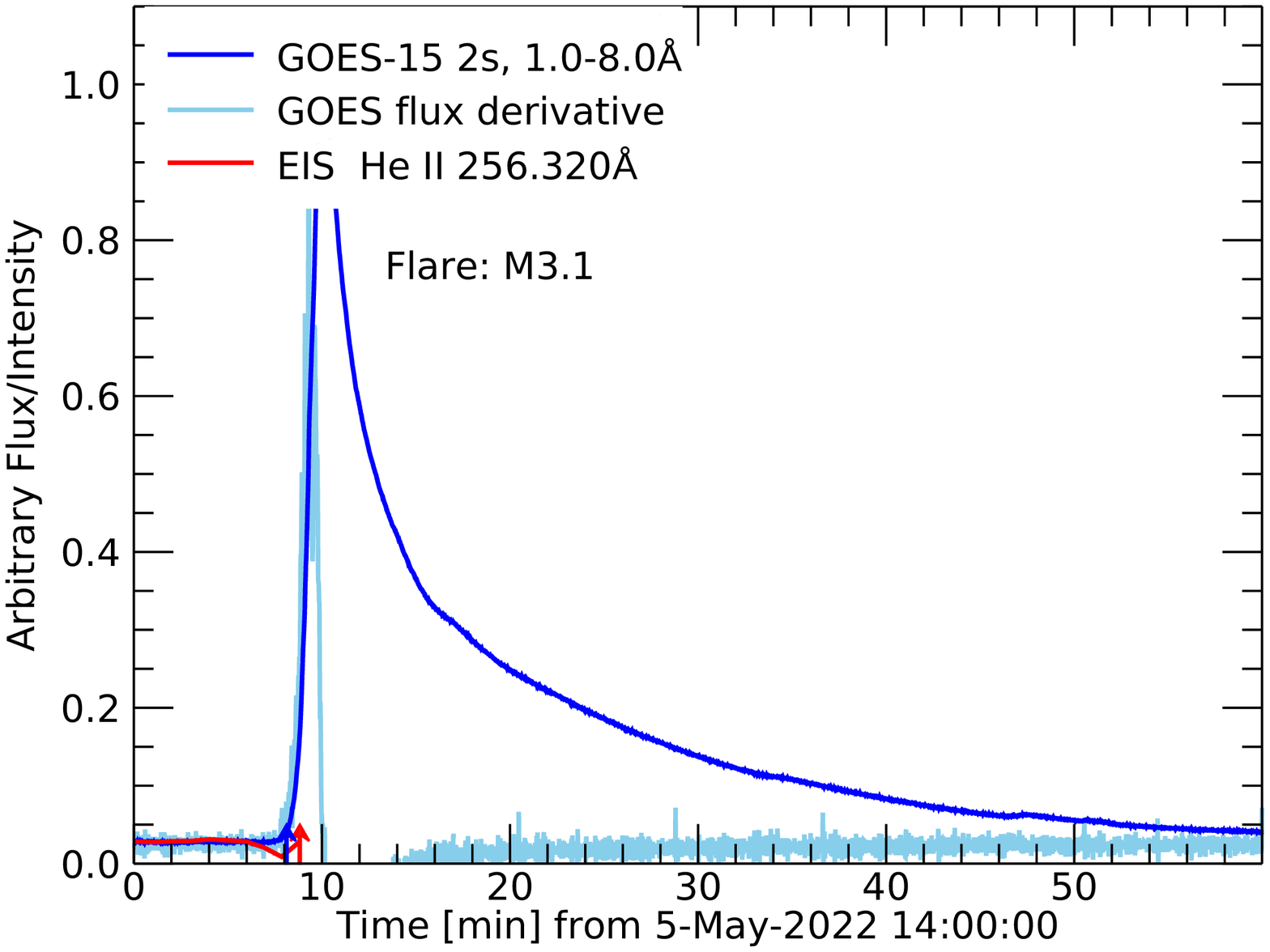}}
\caption{ Same as the bottom row of Figure \ref{fig:1} for the other flares in our sample.
The X-axis labels give the time in minutes from the start time shown in the axis title. The change to 0 in some of the plots indicates one hour. 
}\label{fig:3}
\end{figure*}

Figure \ref{fig:3} shows the results for the remaining 12 flares in our sample. We used the HXR data from RHESSI when available. Sometimes
the data were compromised by changes in attentuator state during the flare, or were affected by eclipses and other issues. In these cases we used the Fermi/GBM
data. On occasion Fermi data were also not available. In these cases we used the derivative of the GOES SXR flux as a proxy for the HXR emission, based on the Neupert
effect\citep{Neupert1968}.
Table \ref{table} gives some relevant details of our flare sample such as the GOES-class, defined GOES start time, measured delay until the triggered response, and flare location.

\begin{figure*}[t!]
\centerline{%
\includegraphics[viewport = 20 10 610 490,scale=0.38]{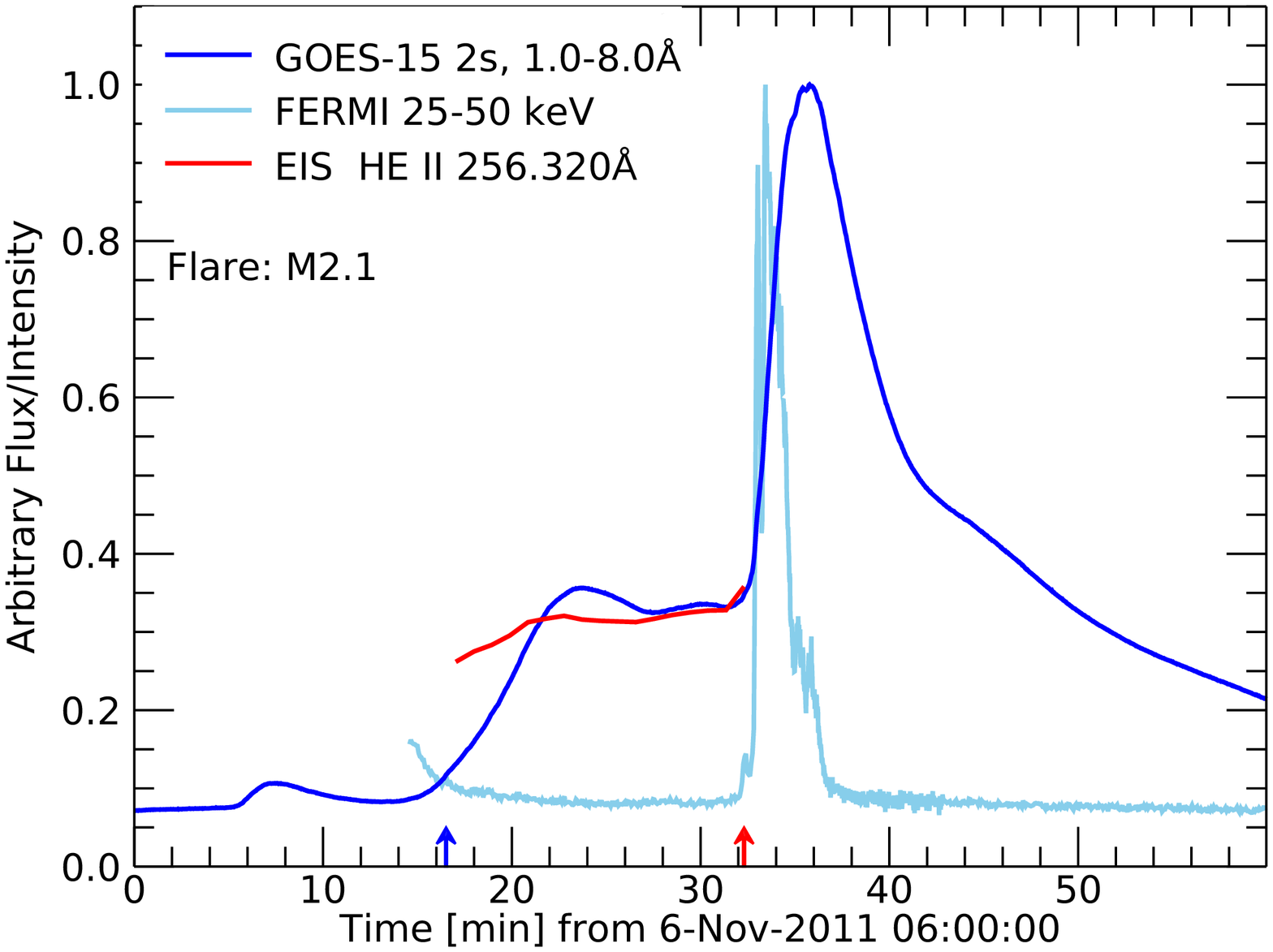}
\includegraphics[viewport = 20 10 610 490,scale=0.38]{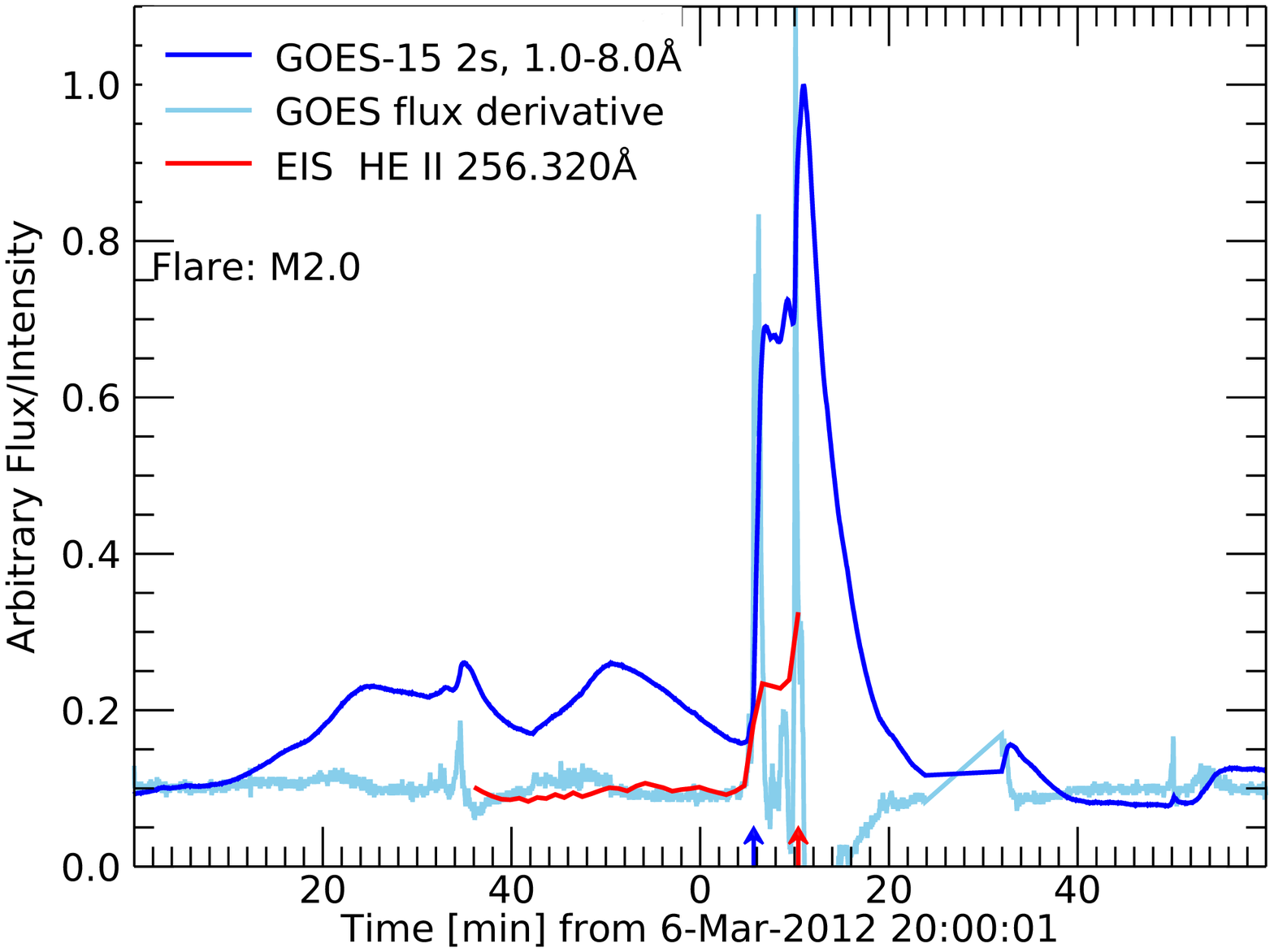}}
\centerline{%
\includegraphics[viewport = 20 10 610 490,scale=0.38]{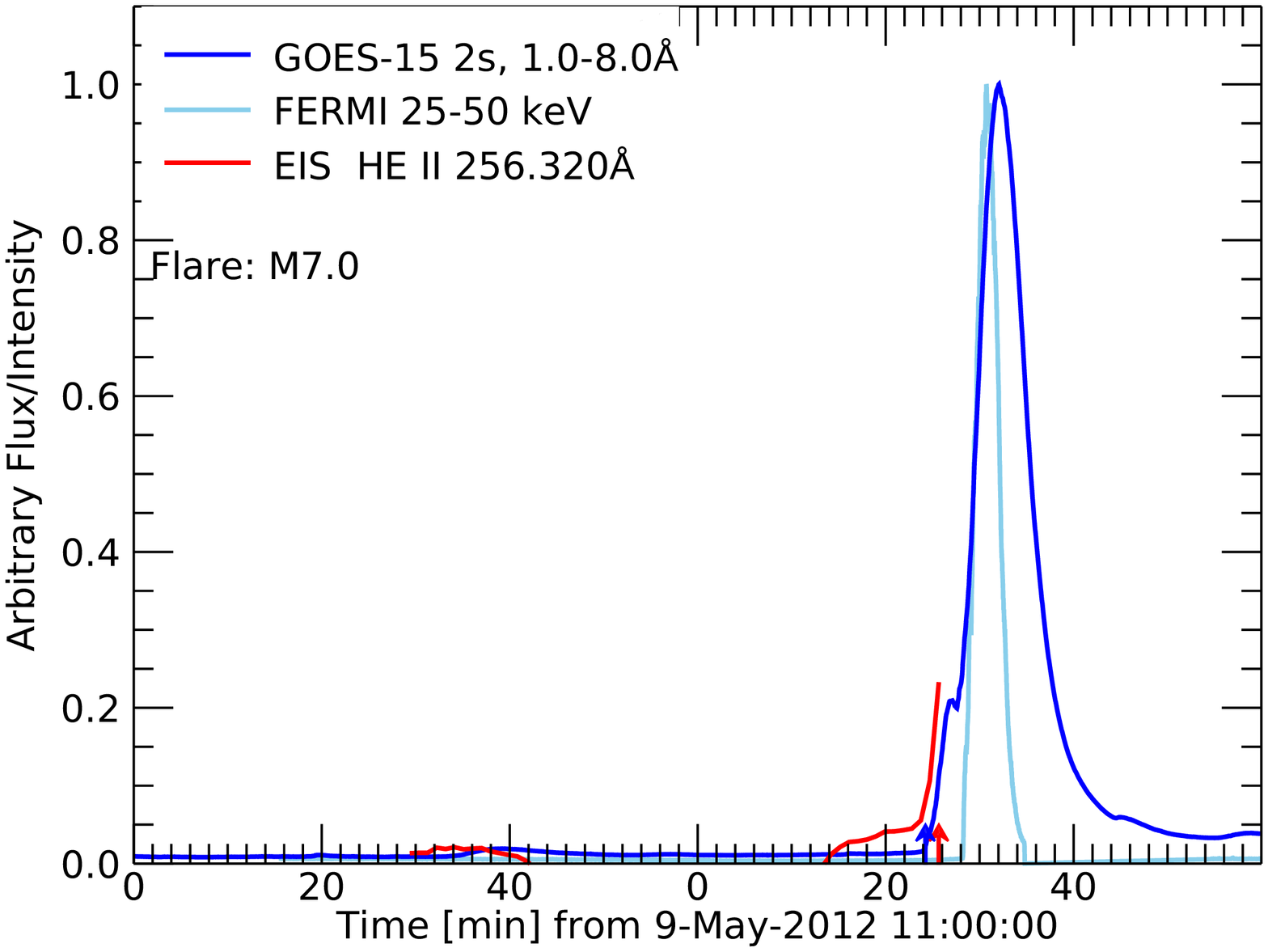}
\includegraphics[viewport = 20 10 610 490,scale=0.38]{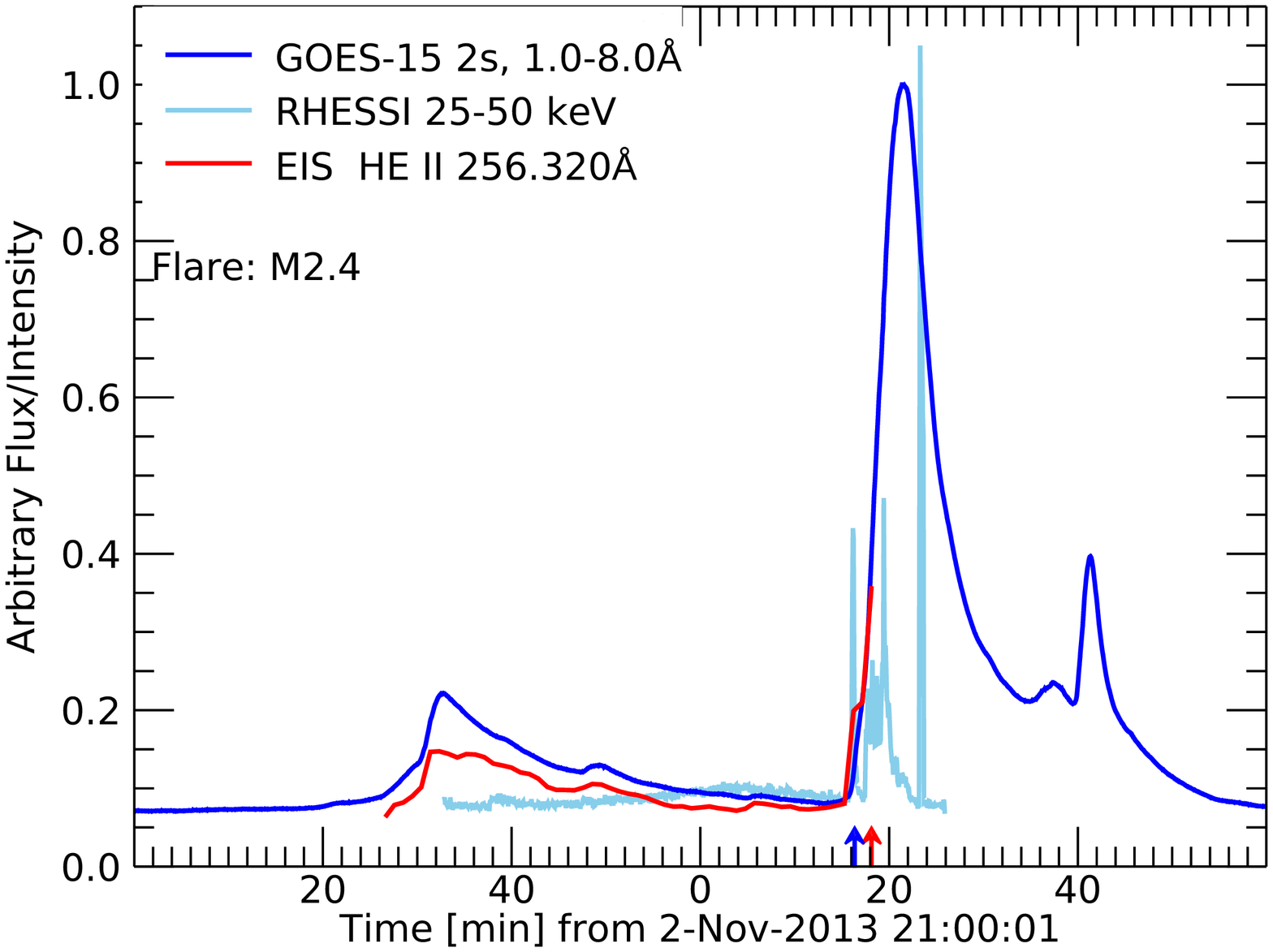}}
\centerline{%
\includegraphics[viewport = 20 10 610 490,scale=0.38]{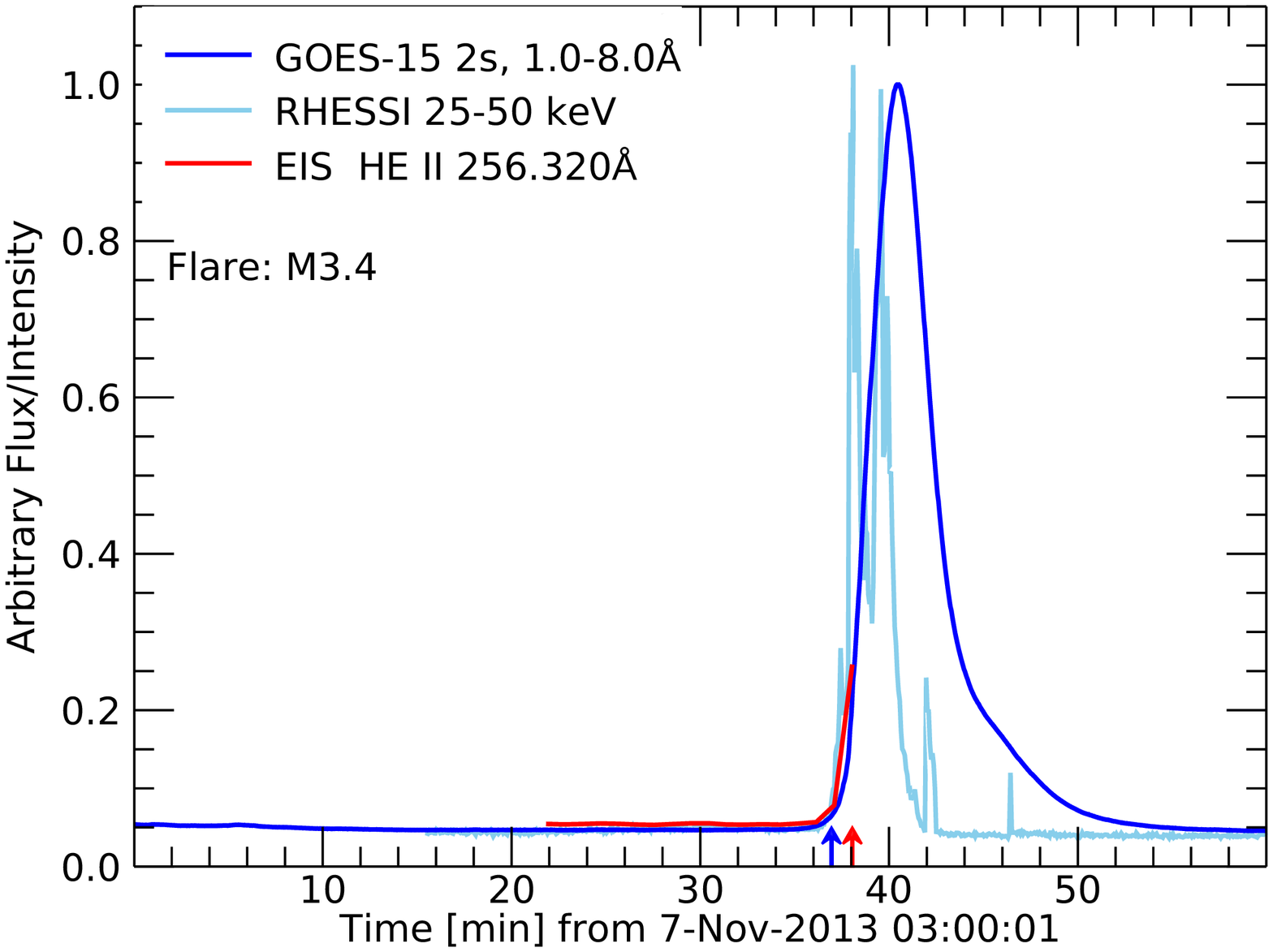}
\includegraphics[viewport = 20 10 610 490,scale=0.38]{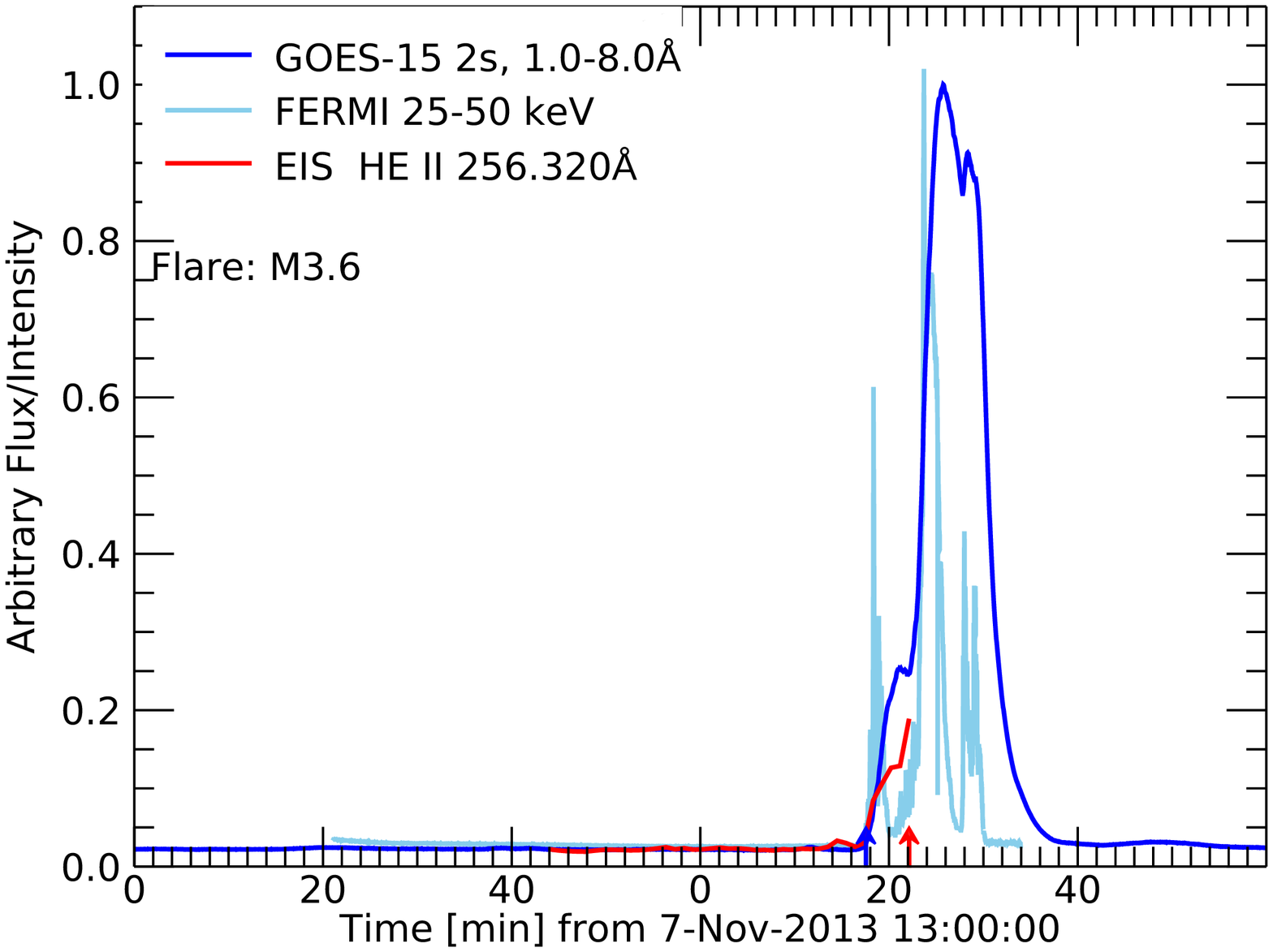}}
\end{figure*}
\newpage
\begin{figure}[h!]
\begin{center}
\includegraphics[width=0.5\linewidth]{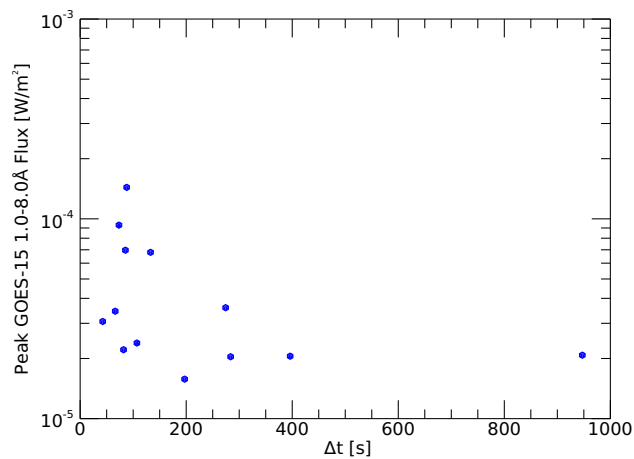}
\end{center}
\caption{ Time delays between the start of the GOES SXR flare and the EIS triggered response plotted as a function of the peak SXR flux. \label{fig:4}}
\end{figure}

Figure \ref{fig:4} summarizes the time-delay between the GOES SXR start time and the EIS triggered response for each flare. 
Generally, the EFT appears to be quite well tuned to the start of the flare and responds promptly, within $\sim$ 2 minutes. There is one outlier case where the time-delay is long (947\,s). On closer inspection, however,
there are extenuating circumstances. This M2.1 flare occurred on 2011, November 6, and is shown in the top left panel of Figure \ref{fig:3}. In this
example, the flare has a double peak at $\sim$06:24\,UT\, and $\sim$06:36\,UT. Our algorithm defines the start time as an increase above the background
at 06:16\,UT, before the first (lower GOES-class) burst. In fact the EFT hunter study did not start until 06:17\,UT, after the flare had begun.         
Of course a flare was occurring so EIS should trigger, but the fact that there is a pedastal and/or decrease in SXR flux before the M2.1 peak apparently fools the
EFT into not detecting the conditions to trigger. When encountering the second peak, however, the EFT apparently does trigger fast (62\,s\, early). 

There is another similar case on 2017, September 6 (lower left panel in Figure \ref{fig:3}). For this flare, the time-delay is 396\,s, which is relatively
long. This flare, however, is triggering in the tail of decreasing SXR flux from an X9.3 event that occurred 7 hours before at $\sim$12\,UT.
This potentially affected the assessment of the background level before the flare.

Excluding these two cases the mean response time for the sample is 130\,s. The uncertainty (standard deviation of the time delays for the sample) is 84\,s. In the best case, the highly impulsive M3.1 flare on 2022, May 5, the
EFT responds in 42\,s.

\section{Discussion}

We assessed the on-orbit performance of the EIS internal flare trigger from a sample of 13 $>$ M-class flares that were successfuly
captured. Excluding two events that were complex to interpret, the mean delay-time between the GOES SXR onset time and the EFT
response was 2 min 10s. We conclude that the EFT, based on intensity thresholding in the He II 256.32\,\AA\, spectral line, is able to react fairly early in the flare evolution. 

Of course there may be scientific studies focused even earlier in the flare impulsive phase, so it makes sense to consider potential
improvements, not just for EIS, but for future missions such as MUSE and Solar-C EUVST. The $\sim$2 min delay suggests that for EIS
the detection thresholds could potentially be reduced, and we showed one example where the flare could have been triggered 1--2\,mins earlier.

When comparing with the HXR data we initially used the GOES SXR derivative as a proxy. \cite{Dennis1993} showed that, following
the Neupert effect,
peaks in the GOES SXR derivative plot corresponded to peaks in the HXR data to within $\pm$20\,s in 88\% of the flares they analyzed.
Our analysis found that the EFT time is closer to the onset
of the increase in HXR, suggesting that this might even be a better trigger criteria. 
This initial conclusion does not look as convincing when we look at the actual HXR data
from RHESSI and Fermi in Figure \ref{fig:3}. 
In several cases the HXR onset follows the SXR rise. This could reflect the specific characteristics of the flares in our sample. \cite{Dennis1993}
did find that the correlation between the SXR time derivative and HXR time profile decreased to 50\% for larger ($>$ X1), gradual events in their study.

Conversely, it would not be practical
to try to implement an on-board algorithm that triggers off of a spacecraft to spacecraft transmission, so an internal running difference of the derivative
of a time-series of intensities from a high temperature line already included in the design wavelength selection (such as Fe XXIV 192.04\,\AA) could be worth investigating
for future missions if chromospheric or transition region coverage is limited.

\section*{Conflict of Interest Statement}

The authors declare that the research was conducted in the absence of any commercial or financial relationships that could be construed as a potential conflict of interest.

\section*{Author Contributions}

Discussions between the authors initiated the study. D.H.B. performed the analysis and wrote the paper. J.W.R. processed the RHESSI and Fermi data. All authors discussed the results and commented on the manuscript. 

\section*{Funding}
The work of D.H.B. and H.P.W. was funded by the NASA Hinode program.

\section*{Acknowledgments}
Hinode is a Japanese mission developed and launched by ISAS/JAXA, with NAOJ 
as domestic partner and NASA and STFC (UK) as international partners. 
It is operated by these agencies in co-operation with ESA and NSC (Norway). 
The AIA images are courtesy of NASA/SDO and the AIA, EVE, and HMI science teams.

\begin{table}[]
\centering
\small
\begin{tabular}{cccc}
\hline
GOES-15 SXR & GOES-15 & EIS & Location \\
Start Time[UT] & Class & $\Delta$t[s] & \\
\hline
6-Nov-2011 06:16:31 & M2.1 & 947.2 & N21E33 \\
6-Mar-2012 21:05:40 & M2.0 & 283.8 & N16E30 \\
9-May-2012 12:24:14 & M7.0 & 85.4 & N13E31 \\
2-Nov-2013 22:16:21 & M2.4 & 107.0 & S12W11 \\
7-Nov-2013 03:36:57 & M3.4 & 66.2 & S14E28 \\
7-Nov-2013 14:17:33 & M3.6 & 274.4 & S13E23 \\
31-Dec-2013 21:48:20 & M9.3 & 73.2 & S16W35 \\
1-Jan-2014 18:42:52 & X1.4 & 87.8 & S14W47 \\
2-Feb-2014 21:27:38 & M1.6 & 197.1 & S10E01 \\
6-Feb-2014 22:59:33 & M2.2 & 81.8 & S15W48 \\
1-Oct-2015 13:06:25 & M6.8 & 132.8 & S23W64 \\
6-Sep-2017 19:22:45 & M2.1 & 396.1 & S08W38 \\
5-May-2022 14:08:07 & M3.1 & 42.5 & S29E64 \\
\hline
\end{tabular}
\caption{GOES-15 start time and class are derived from the 2\,s high cadence 1.0--8.0\,\AA\, data. $\Delta$t is the time delay between the GOES-15 start time and the EFT response. Locations are taken from the Hinode Flare Catalog.}
\label{table}
\end{table}


\begin{thebibliography}{26}
\providecommand{\natexlab}[1]{#1}
\expandafter\ifx\csname urlstyle\endcsname\relax
  \providecommand{\doi}[1]{doi:\discretionary{}{}{}#1}\else
  \providecommand{\doi}{doi:\discretionary{}{}{}\begingroup
  \urlstyle{rm}\Url}\fi
\providecommand{\selectlanguage}[1]{\relax}
\providecommand{\bibAnnoteFile}[1]{%
  \IfFileExists{#1}{\begin{quotation}\noindent\textsc{Key:} #1\\
  \textsc{Annotation:}\ \input{#1}\end{quotation}}{}}
\providecommand{\bibAnnote}[2]{%
  \begin{quotation}\noindent\textsc{Key:} #1\\
  \textsc{Annotation:}\ #2\end{quotation}}

\bibitem[{{Brooks} et~al.(2015){Brooks}, {Ugarte-Urra}, and
  {Warren}}]{Brooks2015}
{Brooks}, D.~H., {Ugarte-Urra}, I., and {Warren}, H.~P. (2015).
\newblock {Full-Sun observations for identifying the source of the slow solar
  wind}.
\newblock \emph{Nature Communications} 6, 5947.
\newblock \doi{10.1038/ncomms6947}
\input{Brooks2015}

\bibitem[{{Bryans} et~al.(2020){Bryans}, {McIntosh}, {Brooks}, and {De
  Pontieu}}]{Bryans2020}
{Bryans}, P., {McIntosh}, S.~W., {Brooks}, D.~H., and {De Pontieu}, B. (2020).
\newblock {Investigating the Chromospheric Footpoints of the Solar Wind}.
\newblock \emph{ApJL} 905, L33.
\newblock \doi{10.3847/2041-8213/abce69}
\input{Bryans2020}

\bibitem[{Culhane et~al.(2007)Culhane, Harra, James, Al-Janabi, Bradley,
  Chaudry et~al.}]{Culhane2007}
Culhane, J.~L., Harra, L.~K., James, A.~M., Al-Janabi, K., Bradley, L.~J.,
  Chaudry, R.~A., et~al. (2007).
\newblock The euv imaging spectrometer for hinode.
\newblock \emph{Sol. Phys.} 243, 19--61.
\newblock \doi{10.1007/s01007-007-0293-1}
\input{Culhane2007}

\bibitem[{{De Pontieu} et~al.(2020){De Pontieu}, {Mart{\'\i}nez-Sykora},
  {Testa}, {Winebarger}, {Daw}, {Hansteen} et~al.}]{DePontieu2020}
{De Pontieu}, B., {Mart{\'\i}nez-Sykora}, J., {Testa}, P., {Winebarger}, A.~R.,
  {Daw}, A., {Hansteen}, V., et~al. (2020).
\newblock {The Multi-slit Approach to Coronal Spectroscopy with the Multi-slit
  Solar Explorer (MUSE)}.
\newblock \emph{ApJ} 888, 3.
\newblock \doi{10.3847/1538-4357/ab5b03}
\input{DePontieu2020}

\bibitem[{{Dennis} and {Zarro}(1993)}]{Dennis1993}
{Dennis}, B.~R. and {Zarro}, D.~M. (1993).
\newblock {The Neupert Effect - what can it Tell up about the Impulsive and
  Gradual Phases of Solar Flares}.
\newblock \emph{Sol. Phys.} 146, 177--190.
\newblock \doi{10.1007/BF00662178}
\input{Dennis1993}

\bibitem[{{Doschek} et~al.(2018){Doschek}, {Warren}, {Harra}, {Culhane},
  {Watanabe}, and {Hara}}]{Doschek2018}
{Doschek}, G.~A., {Warren}, H.~P., {Harra}, L.~K., {Culhane}, J.~L.,
  {Watanabe}, T., and {Hara}, H. (2018).
\newblock {Photospheric and Coronal Abundances in an X8.3 Class Limb Flare}.
\newblock \emph{ApJ} 853, 178.
\newblock \doi{10.3847/1538-4357/aaa4f5}
\input{Doschek2018}

\bibitem[{Freeland and Handy(1998)}]{Freeland1998}
Freeland, S.~L. and Handy, B.~N. (1998).
\newblock Data analysis with the solarsoft system.
\newblock \emph{Sol. Phys.} 182, 497--500.
\newblock \doi{10.1023/A:1005038224881}
\input{Freeland1998}

\bibitem[{Golub et~al.(2007)Golub, DeLuca, Austin, Bookbinder, Caldwell,
  Cheimets et~al.}]{golub2007}
Golub, L., DeLuca, E., Austin, G., Bookbinder, J., Caldwell, D., Cheimets, P.,
  et~al. (2007).
\newblock The x-ray telescope (xrt) for the hinode mission.
\newblock \emph{Sol. Phys.} 243, 63--86.
\newblock \doi{10.1007/s11207-007-0182-1}
\input{golub2007}

\bibitem[{{Harra} et~al.(2017){Harra}, {Hara}, {Doschek}, {Matthews}, {Warren},
  {Culhane} et~al.}]{Harra2017}
{Harra}, L.~K., {Hara}, H., {Doschek}, G.~A., {Matthews}, S., {Warren}, H.,
  {Culhane}, J.~L., et~al. (2017).
\newblock {Measuring Velocities in the Early Stage of an Eruption: Using
  {\textquotedblleft}Overlappogram{\textquotedblright} Data from Hinode EIS}.
\newblock \emph{ApJ} 842, 58.
\newblock \doi{10.3847/1538-4357/aa7411}
\input{Harra2017}

\bibitem[{{Harra} et~al.(2009){Harra}, {Williams}, {Wallace}, {Magara}, {Hara},
  {Tsuneta} et~al.}]{Harra2009}
{Harra}, L.~K., {Williams}, D.~R., {Wallace}, A.~J., {Magara}, T., {Hara}, H.,
  {Tsuneta}, S., et~al. (2009).
\newblock {Coronal Nonthermal Velocity Following Helicity Injection Before an
  X-Class Flare}.
\newblock \emph{ApJL} 691, L99--L102.
\newblock \doi{10.1088/0004-637X/691/2/L99}
\input{Harra2009}

\bibitem[{{Hinode Review Team} et~al.(2019){Hinode Review Team}, {Al-Janabi},
  {Antolin}, {Baker}, {Bellot Rubio}, {Bradley} et~al.}]{Hinode2019}
{Hinode Review Team}, {Al-Janabi}, K., {Antolin}, P., {Baker}, D., {Bellot
  Rubio}, L.~R., {Bradley}, L., et~al. (2019).
\newblock {Achievements of Hinode in the first eleven years}.
\newblock \emph{PASJ} 71, R1.
\newblock \doi{10.1093/pasj/psz084}
\input{Hinode2019}

\bibitem[{{Inglis} et~al.(2021){Inglis}, {Ireland}, {Shih}, and
  {Christe}}]{Inglis2021}
{Inglis}, A.~R., {Ireland}, J., {Shih}, A.~Y., and {Christe}, S.~D. (2021).
\newblock {Evaluating Pointing Strategies for Future Solar Flare Missions}.
\newblock \emph{Sol. Phys.} 296, 153.
\newblock \doi{10.1007/s11207-021-01896-0}
\input{Inglis2021}

\bibitem[{{Jeffrey} et~al.(2018){Jeffrey}, {Fletcher}, {Labrosse}, and
  {Sim{\~o}es}}]{Jeffrey2018}
{Jeffrey}, N.~L.~S., {Fletcher}, L., {Labrosse}, N., and {Sim{\~o}es}, P.~J.~A.
  (2018).
\newblock {The development of lower-atmosphere turbulence early in a solar
  flare}.
\newblock \emph{Science Advances} 4, 2794.
\newblock \doi{10.1126/sciadv.aav2794}
\input{Jeffrey2018}

\bibitem[{{Landi} et~al.(2021){Landi}, {Li}, {Brage}, and {Hutton}}]{Landi2021}
{Landi}, E., {Li}, W., {Brage}, T., and {Hutton}, R. (2021).
\newblock {Hinode/EIS Coronal Magnetic Field Measurements at the Onset of a C2
  Flare}.
\newblock \emph{ApJ} 913, 1.
\newblock \doi{10.3847/1538-4357/abf6d1}
\input{Landi2021}

\bibitem[{{Lemen} et~al.(2012){Lemen}, {Title}, {Akin}, {Boerner}, {Chou},
  {Drake} et~al.}]{Lemen2012}
{Lemen}, J.~R., {Title}, A.~M., {Akin}, D.~J., {Boerner}, P.~F., {Chou}, C.,
  {Drake}, J.~F., et~al. (2012).
\newblock {The Atmospheric Imaging Assembly (AIA) on the Solar Dynamics
  Observatory (SDO)}.
\newblock \emph{Sol. Phys.} 275, 17--40.
\newblock \doi{10.1007/s11207-011-9776-8}
\input{Lemen2012}

\bibitem[{Lin et~al.(2002)Lin, Dennis, Hurford, Smith, Zehnder, Harvey
  et~al.}]{Lin2002}
Lin, R.~P., Dennis, B.~R., Hurford, G.~J., Smith, D.~M., Zehnder, A., Harvey,
  P.~R., et~al. (2002).
\newblock The reuven ramaty high-energy solar spectroscopic imager (rhessi).
\newblock \emph{Sol. Phys.} 210, 3--32.
\newblock \doi{10.1023/A:1022428818870}
\input{Lin2002}

\bibitem[{Meegan et~al.(2009)Meegan, Lichti, Bhat, Bissaldi, Briggs,
  Connaughton et~al.}]{Meegan2009}
Meegan, C., Lichti, G., Bhat, P.~N., Bissaldi, E., Briggs, M.~S., Connaughton,
  V., et~al. (2009).
\newblock The fermi gamma-ray burst monitor.
\newblock \emph{ApJ} 702, 791--804.
\newblock \doi{10.1088/0004-637X/702/1/791}
\input{Meegan2009}

\bibitem[{{Milligan}(2015)}]{Milligan2015}
{Milligan}, R.~O. (2015).
\newblock {Extreme Ultra-Violet Spectroscopy of the Lower Solar Atmosphere
  During Solar Flares (Invited Review)}.
\newblock \emph{Sol. Phys.} 290, 3399--3423.
\newblock \doi{10.1007/s11207-015-0748-2}
\input{Milligan2015}

\bibitem[{{Neupert}(1968)}]{Neupert1968}
{Neupert}, W.~M. (1968).
\newblock {Comparison of Solar X-Ray Line Emission with Microwave Emission
  during Flares}.
\newblock \emph{ApJL} 153, L59.
\newblock \doi{10.1086/180220}
\input{Neupert1968}

\bibitem[{{Pesnell} et~al.(2012){Pesnell}, {Thompson}, and
  {Chamberlin}}]{Pesnell2012}
{Pesnell}, W.~D., {Thompson}, B.~J., and {Chamberlin}, P.~C. (2012).
\newblock {The Solar Dynamics Observatory (SDO)}.
\newblock \emph{Sol. Phys.} 275, 3--15.
\newblock \doi{10.1007/s11207-011-9841-3}
\input{Pesnell2012}

\bibitem[{{Thompson} and {Brekke}(2000)}]{Thompson2000}
{Thompson}, W.~T. and {Brekke}, P. (2000).
\newblock {EUV Full-Sun Imaged Spectral Atlas Using the SOHO Coronal Diagnostic
  Spectrometer}.
\newblock \emph{Sol. Phys.} 195, 45--74.
\newblock \doi{10.1023/A:1005203001242}
\input{Thompson2000}

\bibitem[{{To} et~al.(2021){To}, {Long}, {Baker}, {Brooks}, {van
  Driel-Gesztelyi}, {Laming} et~al.}]{To2021}
{To}, A. S.~H., {Long}, D.~M., {Baker}, D., {Brooks}, D.~H., {van
  Driel-Gesztelyi}, L., {Laming}, J.~M., et~al. (2021).
\newblock {The Evolution of Plasma Composition during a Solar Flare}.
\newblock \emph{ApJ} 911, 86.
\newblock \doi{10.3847/1538-4357/abe85a}
\input{To2021}

\bibitem[{{Tolbert} and {Schwartz}(2020)}]{Tolbert2020}
[Dataset] {Tolbert}, K. and {Schwartz}, R. (2020).
\newblock {OSPEX: Object Spectral Executive}.
\newblock Astrophysics Source Code Library, record ascl:2007.018
\input{Tolbert2020}

\bibitem[{Ugarte-Urra et~al.(2023)Ugarte-Urra, Young, Brooks, Warren,
  De~Pontieu, Bryans et~al.}]{Ugarte2023}
Ugarte-Urra, I., Young, P.~R., Brooks, D.~H., Warren, H.~P., De~Pontieu, B.,
  Bryans, P., et~al. (2023).
\newblock The case for solar full-disk spectral diagnostics: Chromosphere to
  corona.
\newblock \emph{Frontiers in Astronomy and Space Sciences} 9.
\newblock \doi{10.3389/fspas.2022.1064192}
\input{Ugarte2023}

\bibitem[{{Warren} et~al.(2018){Warren}, {Brooks}, {Ugarte-Urra}, {Reep},
  {Crump}, and {Doschek}}]{Warren2018}
{Warren}, H.~P., {Brooks}, D.~H., {Ugarte-Urra}, I., {Reep}, J.~W., {Crump},
  N.~A., and {Doschek}, G.~A. (2018).
\newblock {Spectroscopic Observations of Current Sheet Formation and
  Evolution}.
\newblock \emph{ApJ} 854, 122.
\newblock \doi{10.3847/1538-4357/aaa9b8}
\input{Warren2018}

\bibitem[{Watanabe et~al.(2012)Watanabe, Masuda, and Segawa}]{Watanabe2012}
Watanabe, K., Masuda, S., and Segawa, T. (2012).
\newblock Hinode flare catalogue.
\newblock \emph{Sol. Phys.} 279, 317--322.
\newblock \doi{10.1007/s11207-012-9983-y}
\input{Watanabe2012}

\end{thebibliography}
\end{document}